\documentclass[twocolumn,showpacs,preprintnumbers,amsmath,amssymb]{revtex4}
\usepackage{graphicx}
\usepackage{dcolumn}
\usepackage{bm}
\usepackage{amssymb}
\usepackage{graphicx}
\usepackage{amsmath}
\usepackage{xspace}

\begin{document}

\title{Raman scattering study of electron-doped Pr$_x$Ca$_{1-x}$Fe$_2$As$_2$ superconductors}
\author{A.~P.~Litvinchuk$^{1}$, Bing Lv$^{1}$, and C.~W.~Chu$^{1,2}$}
\affiliation{
$^1$Texas Center for Superconductivity and Department of Physics, University of Houston, TX 77204-5002, USA\\
$^2$Lawrence Berkeley National Laboratory, Berkeley, California 94720, USA }
\date{\today}

\begin{abstract}
Temperature-dependent polarized Raman spectra of electron-doped superconducting
Pr$_x$Ca$_{1-x}$Fe$_2$As$_2$ ($x \approx 0.12$) single crystals are reported.
All four allowed by symmetry even-parity phonons are identified.
Phonon mode of B$_{1g}$ symmetry at 222 cm$^{-1}$, which is associated with the $c$-axis motion of Fe ions,
is found to exhibit an anomalous frequency hardening at low temperatures,
that signals non-vanishing electron-phonon coupling in the superconducting state and implies that the superconducting
gap magnitude $2\Delta_c < 27$meV.
\end{abstract}

\maketitle

Since first reports on superconductivity in iron-arsenide based oxypnictides
LaF$_x$O$_{1-x}$FeAs\cite{ROFA1,ROFA2},
a number of closely related compounds have been discovered \cite{tapp,rotter,sasmal}
and there is a continuing effort in optimizing their superconducting properties and raising
superconductivity transition temperature. It is well understood that, like for other superconductors,
one of the  key factors in achieving this goal is proper material doping.
As far as $Ae$Fe$_2$As$_2$ (where $Ae$ is an alkali earth element) systems is concerned, until
recently the highest superconducting temperature on record was at T$_c$=38~K for the
K$_x$Ba$_{1-x}$Fe$_2$As$_2$ system\cite{rotter,sasmal}, where optimum doping takes place at $x$~$\approx$~0.4.
In that case indirect hole doping of FeAs layers was realized.

Several attempts have been reported to introduce electron doping into $Ae$Fe$_2$As$_2$
system in a search for materials with higher T$_c$'s.
Muraba {\it et al.}\cite{muraba} observed superconductivity up to T$_c$=22~K in the
case of La$^{3+}$ for Sr$^{2+}$ substitution.
Theoretical calculations show clear differences between hole- and electron-doped systems\cite{mazin1}
and predict that superconductivity can sustain higher hole than electron doping\cite{ikeda}.
Despite of this fact it was found recently that the superconducting transition temperature T$_c$
reaches 45~K (under pressure) in the case of La$^{3+}$ for Ca$^{2+}$ substitution \cite{La-doped,qi,saha},
and as high as 49~K for Pr$^{3+}$ indirect doping, as shown by resistive, magnetic and
thermoelectric measurements\cite{lv}. It is unclear, however, whether the superconducting transition
observed is bulk\cite{La-doped} or non-bulk\cite{lv} due to interfacial or filamentary superconductivity.
Interfacial or filamentary superconductivity has been suggested to lead to an enhanced T$_c$\cite{allender}.

In this communication we report the results of a Raman scattering study of Pr$^{3+}$ doped
CaFe$_2$As$_2$ single crystals. Even though phonons themselves and the strength of electron-phonon interaction
are not sufficient to explain the rather high T$_c$ in this class of materials\cite{isotop},
we will focus on the phonon spectra and lattice vibrations which,
due to their coupling to the electronic states, may shed light on the origin of superconductivity
and the details of the interplay between lattice, charge, and spin degrees of freedom.
The samples studied here were CaFe$_2$As$_2$ single crystals doped with Pr-doping at the level $x~\approx$~0.12.
They showed two superconducting transitions with T$_c$=49~K and 21~K. Preparation technique and
characterization of single crystals by X-ray diffraction, transport, and magnetic susceptibility
measurements have been reported in Ref.~\onlinecite{lv}.

Raman scattering measurements were performed with a triple Horiba Jobin Yvon T64000
spectrometer, equipped with an optical microscope and liquid-nitrogen-cooled CCD
detector. Samples were mounted on the cold finger of a helium flow optical
cryostat and the temperature was controlled within 0.1~K.
He-Ne laser ($\lambda_{las}=638.2$nm) was used as the
excitation source and the power density did not exceed $10^4$~W/cm$^2$ in
order to minimize heating of the sample.

\begin{figure}[hb]
\includegraphics[width=7.5cm]{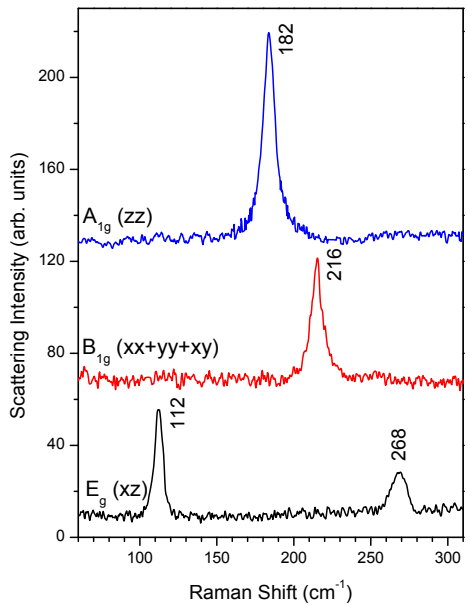}
\caption{(Color online) Raman spectra of Pr$_x$Ca$_{1-x}$Fe$_2$As$_2$ for
selected scattering geometries taken at room temperature.
Allowed mode symmetries are also listed.}
\end{figure}

$Ae$Fe$_2$As$_2$ compounds crystallize in the tetragonal ThCr$_2$Si$_2$ type
structure\cite{rozsa} within the space group I4/mmm and their vibrational spectra
are well understood\cite{litvin08}. Model calculations (density functional theory
as well as shell model) reproduce experimental phonon frequencies rather well and provide
information on displacement patterns for lattice eigenmodes. The alkaline earth element
does not contribute to any of the Raman active modes as it occupies
a centrosymmetric position within the lattice. Fe and As ions contribute B$_{1g}$ and A$_{1g}$ modes,
respectively, which correspond to atomic displacements along $c$-axis.
$ab$-plane displacements of Fe and As produce two E$_g$ modes, which involve motion of both ions
and are therefore strongly mixed \cite{litvin08}.
The experimental frequencies of the Raman-active modes (at room temperature) are 114, 182, 204 and 264 cm$^{-1}$
(E$_g$, A$_{1g}$, B$_{1g}$, E$_g$, respectively) for undoped SrFe$_2$As$_2$;
124, 209, 264 cm$^{-1}$ (the A$_{1g}$ mode was not observed from the $ab$-surfaces of the crystals)
for BaFe$_2$As$_2$\cite{Ba},
and 189, 211 cm$^{-1}$ (A$_{1g}$ and B$_{1g}$) for CaFe$_2$As$_2$\cite{Ca}.
Thus, for the reason given above,  the mode frequencies do not vary strongly upon replacement of
the alkaline earth element in the structure.

Fig.~1 shows room temperature Raman scattering spectra of Pr-doped CaFe$_2$As$_2$ for different
scattering geometries. The spectra were taken from the edge of a thin crystal with
well developed $(a-b)$-surface, so that the light was propagating in the $ab$-plane.
This geometry allowed us to measure spectra in the back-scattering configuration
with the light polarized either along the $z$-axis or perpendicular to it.
Knowing that the A$_{1g}$ phonon is strong in the (zz) scattering configuration,\cite{litvin08}
we readily identified all four allowed by symmetry modes in Pr$_x$Ca$_{1-x}$Fe$_2$As$_2$:
112, 182, 216, and 268 cm$^{-1}$ (E$_g$, A$_{1g}$, B$_{1g}$, E$_g$, respectively).

Further, we performed temperature-dependence measurements using a crystal,
which allowed to simultaneously observe A$_{1g}$ and B$_{1g}$ modes. Due to the weak scattering
efficiency acquisition time of 60-100 minutes per spectrum was required. Spectra for selected
temperatures are displayed in Fig.~2, and the results of their analysis are summarized in Fig.~3.

\begin{figure}[]
\includegraphics[width=7.5cm]{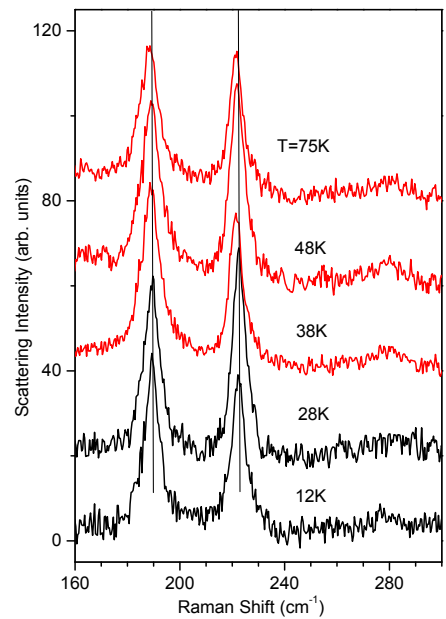}
\caption{(Color online)
Temperature dependent Raman scattering spectra of Pr$_x$Ca$_{1-x}$Fe$_2$As$_2$.
Vertical lines mark the low-temperature position of the lines.}
\end{figure}
\begin{figure}[]
\includegraphics[width=\linewidth]{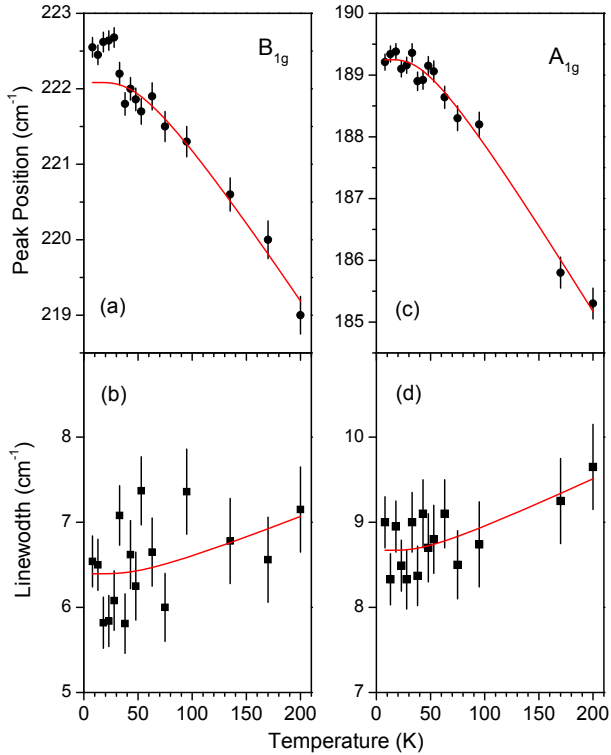}
\caption{(Color online)
Position (a,c) and linewidth (b,d) of A$_{1g}$ and B$_{1g}$ modes as a function
of temperature. Solid lines show the expected behavior due to anharmonic
phonon decay (see text). Note anomalous B$_{1g}$ mode frequency hardening
at low temperatures.}
\end{figure}

The temperature dependence of the linewidth (full width at half-maximum) of the modes appears to be very unusual.
Each of the modes exhibits changes of only about 1 cm$^{-1}$ between 5 and 200~K,
and the low-temperature values of the width are 6.4 and 8.7 cm$^{-1}$ (Fig.~3 (b) and (d) for B$_{1g}$ and A$_{1g}$ modes,
respectively). It implies that, apart from conventional anharmonic phonon decay processes, which typically govern
line broadening as a function of temperature for insulating and semiconducting materials, there are some other
processes involved.

Indeed, phonon anharmonicity leads to the following temperature dependence of the linewidth $\Gamma$ and mode
frequency $\omega$\cite{ipatova,maradudin}:

\begin{align}
\Gamma(T) = \gamma + \Gamma_0(1+{\displaystyle\frac{2}{e^{\hbar\omega_0/{2k_BT}}-1}}), \\
\omega(T) = \omega_0 - C(1+{\displaystyle\frac{2}{e^{\hbar\omega_0/{2k_BT}}-1}}),
\end{align}

\noindent where $\hbar\omega_0$ is the phonon frequency, $k_B$ - Boltzmann constant, and $C$ is a constant.
For insulators and semiconductors $\Gamma_0$ is the only parameter needed to adequately describe the
temperature dependence of the linewidth\cite{balkan} (so that $\gamma = 0)$.
In case of crystal with defects (impurities, structural imperfections etc.)
an additional temperature-independent component of the linewidth $\gamma$ might be required.
Naturally, phonon interactions with other excitations (electrons in conducting materials, e.g.),
which open additional channels for phonon decay, will influence the phonon lifetime and contribute to the phonon
linewidth $\Gamma(T)$\cite{cardona}. This is in fact the phenomenon, that could be used to monitor features
of electronic spectra such as gap opening in superconductors\cite{zey,friedl,litvin91}
or spin excitations of superconductors and/or magnetically ordered systems\cite{litvin92,coo2,spph}.

Solid lines in Fig. 3 (b) and (d) are plotted following Eg.~1.
The fit of the temperature-dependent linewidth yields $\gamma \gg \Gamma_0$ for both B$_{1g}$ (6.0 vs. 0.4 cm$^{-1}$)
and A$_{1g}$ (8.3 vs. 0.4 cm$^{-1}$) modes. Small values of $\Gamma_0$ with respect to the $\gamma$ clearly signal
that lattice anharmonicity is not a dominant mechanism of the phonon decay in the material under investigation.
Thus, the phonon coupling to electrons in the conducting Pr$_x$Ca$_{1-x}$Fe$_2$As$_2$ governs the T-dependence
of the linewidth over the whole temperature range.
We also note that the observed for Pr-doped CaFe$_2$As$_2$ phonon linewidths, which are shown in Fig.~3,
are larger compared to those reported by Choi et al. in Ref.\onlinecite{Ca} for undoped parent CaFe$_2$As$_2$ compound
(about 4.2 and 4.5 cm$^{-1}$ for A$_{1g}$ and B$_{1g}$ modes, respectively). This could be due to modification
by Pr-doping of the electronic states, to which phonon couple, but also, at least partly, due to substitutional disorder.

Next, using Eg.~2, the predicted anharmonic behavior of mode frequencies is shown by solid lines in Fig. 3 (a,c).
It well describes the temperature dependence of A$_{1g}$ mode.
The B$_{1g}$ mode at 222 cm$^{-1}$ exhibits, however, clear deviation from the expected behavior below 38~K.
This implies redistribution upon cooling of electronic states, to which this mode couples, and is consequence
of a gap opening. Phonon hardening implies that the mode frequency exceeds the gap magnitude 2$\Delta_c$.\cite{zey,friedl}

The fact that B$_{1g}$ phonon mode, but not A$_{1g}$ mode, exhibits specific features upon entering
the superconducting state in electron-doped Pr$_x$Ca$_{1-x}$Fe$_2$As$_2$ is not surprising in view of recent
findings
in theoretical band structure calculations and experimental analysis of symmetry-dependent electron-phonon
coupling and electronic inelastic light scattering.\cite{muschler,mazin2}
Indeed, the A$_{1g}$ scattering channel is shown to probe Brillouin center hole Fermi sheets,
B$_{2g}$ (E$_g$ in the tetragonal notations) is maximal at the edges of the Brillouin zone ($\pi/2,\pi/2$),
where there are no Fermi sheets, while B$_{1g}$ channel couples to the {\it electronic} pockets of the Fermi surface.
That could be the reason for the observed B$_{1g}$-symmetry phonon anomalies in an electron-doped
Pr$_x$Ca$_{1-x}$Fe$_2$As$_2$ superconductors, which where not detected earlier in
the hole-doped K$_x$Sr$_{1-x}$Fe$_2$As$_2$\cite{litvin08}.

In conclusion, all four Raman active phonons in  Pr$_x$Ca$_{1-x}$Fe$_2$As$_2$ ($x \approx$ 0.12)
have been observed experimentally.  Anomalous B$_{1g}$ mode hardening is observed upon entering
the superconducting state, which is associated with the opening of the superconducting gap
below the phonon mode frequency (222 cm$^{-1}$), so that the gap magnitude $2\Delta_c < 27$ meV.

\acknowledgements This work is supported in part by the T.L.L.
Temple Foundation, the J.J. and R.~Moores Endowment, the State of
Texas through TCSUH, the USAF Office of Scientific Research, and
the LBNL through USDOE. Critical reading of the manuscript by M.N.~Iliev is greatly appreciated.

\end{document}